\documentclass[fleqn,usenatbib]{mnras}

\usepackage{newtxtext,newtxmath}

\usepackage[T1]{fontenc}

\DeclareRobustCommand{\VAN}[3]{#2}
\let\VANthebibliography\thebibliography
\def\thebibliography{\DeclareRobustCommand{\VAN}[3]{##3}\VANthebibliography}

\usepackage{graphicx}	%
\usepackage{amsmath}	%
\usepackage{xcolor}
\usepackage{soul}

\title[Simulation vs. Observation with Anomaly Detection]{Quantitatively rating galaxy simulations against real observations with anomaly detection}

\author[Z. Jin et al.]{
Zehao Jin$^{1,2}$,\thanks{E-mail: zj448@nyu.edu}
Andrea V. Macciò$^{1,2,3}$,
Nicholas Faucher$^{4}$,
Mario Pasquato$^{1,2,5,6,7}$,
Tobias Buck$^{8,9}$,
\newauthor{Keri L. Dixon$^{1,2}$,
Nikhil Arora$^{1,2,10}$,    
Marvin Blank$^{1,2}$,
Pavle Vulanović$^{1,2}$}
\\
$^{1}$New York University Abu Dhabi, PO Box 129188, Saadiyat Island, Abu Dhabi, United Arab Emirates\\
$^{2}$Center for Astrophysics and Space Science (CASS), New York University Abu Dhabi\\
$^{3}$Max-Planck-Institut f\"ur Astronomie, K\"onigstuhl 17, 69117 Heidelberg, Germany\\
$^{4}$Center for Cosmology and Particle Physics, Department of Physics, New York University, 726 Broadway, New York, New York 10003, USA\\
$^{5}$Physics and Astronomy Department Galileo Galilei, University of Padova, Vicolo dell’Osservatorio 3, I–35122, Padova\\
$^{6}$ D\`epartement de Physique, Universit\`e de Montr\`eal, 1375 Ave. Thérèse-Lavoie-Roux, Montréal, QC H2V 0B3, Canada\\
$^{7}$ Mila - Quebec Artificial Intelligence Institute, 6666 Rue Saint-Urbain, Montréal, QC H2S3H1, Canada\\
$^{8}$Universit\"at Heidelberg, Interdisziplin\"ares Zentrum f\"ur Wissenschaftliches Rechnen, Im Neuenheimer Feld 205, \\ D-69120 Heidelberg, Germany\\
$^{9}$Universit\"at Heidelberg, Zentrum f\"ur Astronomie, Institut f\"ur Theoretische Astrophysik, Albert-Ueberle-Straße 2, \\ D-69120 Heidelberg, Germany\\
$^{10}$Department of Physics, Engineering Physics \& Astronomy, Queen's University, Kingston, ON K7L 3N6, Canada\\
}

\date{Accepted 2024 February 19. Received 2024 February 13; in original form 2023 May 15}

\pubyear{2024}

\begin{document}
\label{firstpage}
\pagerange{\pageref{firstpage}--\pageref{lastpage}}
\maketitle

\begin{abstract}
Cosmological galaxy formation simulations are powerful tools to understand the complex processes that govern the formation and evolution of galaxies. However, evaluating the realism of these simulations remains a challenge. The two common approaches for evaluating galaxy simulations is either through scaling relations based on a few key physical galaxy properties, or through a set of pre-defined morphological parameters based on galaxy images. This paper proposes a novel image-based method for evaluating the quality of galaxy simulations using unsupervised deep learning anomaly detection techniques. By comparing full galaxy images, our approach can identify and quantify discrepancies between simulated and observed galaxies. As a demonstration, we apply this method to SDSS imaging and NIHAO simulations with different physics models, parameters, and resolution. We further compare the metric of our method to scaling relations as well as morphological parameters. We show that anomaly detection is able to capture similarities and differences between real and simulated objects that scaling relations and morphological parameters are unable to cover, thus indeed providing a new point of view to validate and calibrate cosmological simulations against observed data.
\end{abstract}

\begin{keywords}
galaxies: formation, galaxies: evolution, methods: data analysis,
methods: numerical, methods: statistical, quasars: supermassive black holes
\end{keywords}

\section{Introduction}
Cosmological simulations are one of the most powerful tool to test our standing of Universe \citep{2020NatRP...2...42V}. Thanks to the blooming development in computational power, high resolution hydrodynamic simulations are able to model galaxy formation and evolution, tracking both dark matter and baryons across cosmic time. Recent simulations are quite successful in reproducing a wide range of galaxy properties such as stellar mass, rotation curves, chemical abundances, colors, and scaling relations \citep[e.g.][]{2014MNRAS.444.1518V,10.1093/mnras/sts028,10.1093/mnras/stw2265,2015MNRAS.446..521S,Wang_2015,10.1093/mnras/stx1160,10.1093/mnras/stx458,2018MNRAS.473.4077P,10.1093/mnras/stx3040,10.1093/mnras/stz3289}.

Examining the agreement between real observations and simulations is the best way to assess the quality of simulations, and is crucial to test and optimize the physics modelled in simulations. Generally, the comparison between the distribution of galaxy properties retrieved from simulations and observations served as a diagnostic tool.
In a multi-dimensional parameter space, the diagnostic is performed through galaxy scaling relations. For example, the Tully-Fisher relation \citep{1977A&A....54..661T}, the size mass relation \citep{2007ApJ...671..203C}, the stellar mass-halo mass relation \citep{Moster_2010,Moster2018}, and the mass metallicity relation \citep{2004ApJ...613..898T,10.1111/j.1365-2966.2005.09321.x,2015ApJ...810...56K}, to name a few. The agreement between these scaling relations in real galaxies and in simulated galaxies is often used as a metric to tell if a simulation is `good'.

Although scaling relations are one of the most commonly used way to assess the validity of cosmological simulations, they `zip' billions of data points generated by a simulation down to a set of a few numbers to compare against the same few numbers also distilled from huge amount of data coming from observations. Inevitably, lots of information contained in both simulations and observations is lost during this process. As a result, sometimes a simulation can fit one set of observed scaling relations, while departs from another set of scaling relations at the same time. Even more challenging is the difficulty in quantitatively determining which set of scaling relations holds more significance when contradictions between them arise. It is then natural to look for  alternative ways for data-model comparison in order to take full advantage of the large (spatial) resolution recently attained by both.

Indeed, great effort has been devoted to the analysis of galaxy images, with various metrics and statistics, both parametric ones such as Sérsic parameters \citep{Sersic_1963}, and non-parametric ones such as Gini coefficient, $M_{20}$, bulge statistics \citep[Gini-$M_{20}$ system,][]{Lotz_2004}, concentration, asymmetry, smoothness \citep[CAS statistics,][]{Conselice_2003}, multimode, intensity, and deviation \citep[MID statistics,][]{Freeman_2013}. The comparison between the distribution of these image-based parameters in mock images to that in real images has now started to serve as one of the crucial tools in the calibration for modern simulations \citep{Snyder_2015,Bottrell_2017a,Bottrell_2017b,Rodriguez-Gomez_2019,Bignone_2020,deGraaff_2022}. Over the years, although more and more morphological parameters are being proposed and improving our understanding of galaxy images, it is challenging to fully exploit the parameter space to characterize a galaxy image through a supervised and human-driven way.

Image recognition and generation has been one of the biggest highlights in the field of deep learning, from the early attempts to distinguish cats and dogs, to artificial intelligence (AI) face generators, and very recently to ChatGPT's brother, an incredible drawing AI called DALL-E. Astrophysicists have also tried to apply machine learning techniques
to attack their science problems, especially those related to galaxy images \citep[e.g.][]{10.1093/mnras/stv632,Storey_Fisher_2021,2021arXiv211101154B,10.1093/mnras/sty1022,2021MNRAS.506..150B,Cheng2021, Smith2022MNRAS.511.1808S}. A very recent work by \citep{Tohill_2023} further shows unsupervised machining trained on galaxy images can encode images into features that among which some are correlated to known morphological parameters, such as Sérsic index, asymmetry and concentration (see \citealt{Tohill_2023} Table 3).

Particularly, deep learning anomaly detection algorithms have the potential to be very powerful when comparing simulated and real galaxy images. In such studies, real observed galaxy images are treated as `normal' images and a neural network will assign `anomaly scores' to simulated galaxy images which quantifies how realistic these simulated images are. \cite{10.1093/mnras/staa1647} used Wasserstein generative adversarial network (WGANs) \citep{wgan} to find outliers in Horizon-AGN simulation \citep{2014MNRAS.444.1453D}, with H-band CANDELS \citep{2011ApJS..197...35G,2011ApJS..197...36K} images as `norm', and WGAN loss as anomaly score. \cite{2021MNRAS.501.4359Z} further improve the performance of anomaly detection algorithm by combining the output of two separate PixelCNN \citep{https://doi.org/10.48550/arxiv.1601.06759} networks to generate pixel-wise anomaly score without sky background contamination. In their work, the Illustris Simulation \citep{2014MNRAS.444.1518V} and IllustrisTNG \citep{10.1093/mnras/stx3040} were compared to $r$-band Sloan Digital Sky Survey (SDSS) images and disagreement in small-scale morphological details are spotted. In this work, we will utilize GANomaly \citep{akcay2018ganomaly}, featuring a encoder-decoder-encoder generator structure and a better defined anomaly score, to compare NIHAO (Numerical Investigation of Hundred Astrophysical Objects) simulations \citep{Wang_2015} to tri-color SDSS RGB ($i$-$r$-$g$ band) galaxy images. GANomaly is a straightforward, concise and effective way to derive anomaly scores that are only related to galaxy features but not to background noise, while at the same time locating the anomaly\footnote{The model is publicly available at \url{https://github.com/ZehaoJin/Rate-galaxy-simulation-with-Anomaly-Detection}}.

This paper is organized as follows: In Section \ref{sec:simulation} we will introduce different sets of NIHAO simulations that will later be rated by GANomaly. In Section \ref{sec:observation} the training set used in this work, the SDSS galaxy images will be reviewed. Section \ref{sec:GANomaly} outlines how GANomaly and anomaly score works. Section \ref{sec:results}, \ref{sec:scalingrelations}, \ref{sec:morphology} and \ref{sec:background} present the main results of this paper, the comparison to scaling relations and morphological parameters, as well as additional discussion and interpretation on anomaly scores. We conclude in Section \ref{sec:conclusion} and explore the feature space behind GANomaly in the Appendix \ref{sec:latent} by trying to interpret the latent space with principal component analysis (PCA) and attach physical meaning to it.

\section{simulation}
\label{sec:simulation}
We make use of the ``vanilla'' version of the NIHAO simulation (hereafter `NIHAO NoAGN') and its variations NIHAO AGN, NIHAO n80, and NIHAO UHD to make a comparison across different galaxy formation physics models, parameters and resolutions. These simulated galaxies are further mock observed into SDSS-style RGB images. Finally, mock observed images are rated by GANomaly for their anomaly scores. Details on NIHAO simulations and the mock observation scheme are presented below.

\subsection{NIHAO NoAGN}
The NIHAO (Numerical Investigation of Hundred Astrophysical Objects) simulation \citep{Wang_2015,2019MNRAS.487.5476B} is a suite of hydrodynamical cosmological zoom-in simulations powered by the \textsc{\small GASOLINE2} code \citep{2017MNRAS.471.2357W}. NIHAO adopts a flat $\Lambda$CDM cosmology and parameters from the Planck satellite results \citep{2014A&A...571A..16P}. NIHAO includes Compton cooling, photoionization from the ultraviolet background following \cite{2012ApJ...746..125H}, star formation and feedback from supernovae \citep{10.1111/j.1365-2966.2006.11097.x} and massive stars \citep{10.1093/mnras/sts028}, metal cooling, and chemical enrichment. A series of prior work has proven NIHAO simulated galaxies reproduce galaxy scaling relations very well, including the Stellar Halo-Mass relation \citep{Wang_2015}, the disc gas mass and disc size relation \citep{10.1093/mnrasl/slw147}, the Tully-Fisher relation \citep{10.1093/mnras/stx458}, the diversity of galaxy rotation curves \citep{10.1093/mnras/stx2660}, and the mass-metallicity relation \citep{2021MNRAS.508.3365B}. 

We refer to this basic version (detailed in \cite{Wang_2015}) of NIHAO simulations as `NIHAO NoAGN'. NIHAO NoAGN is the basis of other variations of NIHAO described in the subsections below.

\subsection{NIHAO AGN}
As named, `NIHAO NoAGN' does not contain active galactic nuclei (AGN) physics. Since it is widely accepted that black hole feedback is crucial in quenching of elliptical galaxies \citep[e.g.][]{Croton2006,2015MNRAS.453.2447D} \cite{2019MNRAS.487.5476B} introduced black hole formation, accretion and feedback to the NIHAO project. In NIHAO AGN, a black hole is seeded when the central halo exceeds a certain mass threshold and then follows the accretion \citep{1952MNRAS.112..195B} and feedback model introduced by \cite{2005MNRAS.361..776S}, one of the most widely used and thus tested models. More details on the AGN implementation in NIHAO can be found in \cite{2019MNRAS.487.5476B}, as well as in \cite{Waterval_2022} for a nice summary.

Practically, NIHAO AGN is a re-run of NIHAO NoAGN, with the exact same initial conditions, parameters, and physics, except the additional AGN implementation, thus providing the AGN counterpart of all vanilla NIHAO galaxies. It is ideal to test the effect of the implemented AGN model by comparing AGN and NoAGN counterparts with scaling relations\citep{Frosst2022, Waterval_2022}, or now even better, with our anomaly scores.

\subsection{NIHAO n80}
\label{sec:n80}
Galaxy formation involves a huge dynamical range, from molecular clouds to large scale environment, making it impossible to fully resolve some of the key processes. Effective models, often with parameters and thresholds, are usually adopted in cosmological numerical simulations to resolve this sub-resolution problem \citep{2003MNRAS.339..289S}. For example, star formation is usually modeled by a density threshold $n$, in particles per $\textup{cm}^3$. Gas particles will start to turn into star particles, i.e. form stars, only when this threshold is reached. Although in a real Universe the `expected' $n$ is above $10^5\ \textup{cm}^{-3}$ \citep{2007ARA&A..45..565M}, such density is out of reach even for the highest resolution simulations of spiral galaxy, as \cite{2020NatRP...2...42V} reviewed. In fact, current leading cosmological simulations tends to use $n$ around $0.1-1\ \textup{cm}^{-3}$ in the lower end, such as EAGLE \citep{10.1093/mnras/stu2058}, Illustris \citep{2014MNRAS.444.1518V}, IllustrisTNG \citep{10.1093/mnras/stx3040}, and around $10-100\ \textup{cm}^{-3}$ in the higher end, such as \cite{2010Natur.463..203G}, FIRE \citep{2015MNRAS.454.2092O}, \cite{10.1093/mnras/stv1699}, VINTERGATAN \cite{Agertz2021} and NIHAO \citep{Wang_2015}. The exact value of $n$ is usually tuned by galaxy scaling relations.

All other NIHAO simulations described in this work, NIHAO NoAGN, NIHAO AGN, and NIHAO UHD, use $n=10\ \textup{cm}^{-3}$. To explore other values of $n$, \cite{Macci__2022} produces a few re-run of NIHAO NoAGN galaxies with $n=80\ \textup{cm}^{-3}$. To clarify, these NIHAO n80 galaxies do not include AGN physics, have the same (mass) resolution as NIHAO NoAGN and NIHAO AGN, but have $n$ set to 80 $\textup{cm}^{-3}$ instead of 10 $\textup{cm}^{-3}$. For an extensive study of the impact of the star formation threshold on the properties of NIHAO galaxies see \cite{Dutton2019,Buck2019b,Dutton2020}.

\subsection{NIHAO UHD}
NIHAO NoAGN already has quite good resolution: dark matter particle mass from $m_{\textup{dm}}=3.4\times10^3 \textup{M}_\odot$ for dwarf galaxies to $m_{\textup{dm}}=1.4\times10^7 \textup{M}_\odot$ for the most massive galaxies. The ratio between dark and gas particle masses is initially the same as the cosmological dark/baryon mass ratio, $\Omega_{\textup{dm}}/\Omega_{\textup{b}}\approx 5.48$. The gas and star particle force softenings are set to be approximately 2.34 times smaller than those of the dark matter particles \citep{Wang_2015}. Additionally, a few Milky-way-like galaxies are selected to be re-simulated at even higher resolution ($m_{\textup{dm}}\approx10^5 \textup{M}_\odot$). \cite{Buck2020} introduces the NIHAO UHD (Ultra High Definition) suite, which contains higher resolution counterparts of six NIHAO NoAGN galaxies, with the same initial conditions, parameters ($n=10\ \textup{cm}^{-3}$), and physics (no AGN). Those galaxies demonstrate the excellent convergence of the NIHAO simulations and show good agreement between the satellite mass function of MW and M31 \citep{Buck2019} and MW bulge properties \citep{Buck2018,Buck2019c}.

\subsection{Mock observation images}
\label{sec:mockobs}
In this work, we use 77 NIHAO NoAGN galaxies, 77 NIHAO AGN galaxies, 12 NIHAO n80 galaxies, and 6 NIHAO UHD galaxies with each galaxy projected along 20 randomly oriented axes. These galaxy simulations are further mock observed in SDSS $i$-$r$-$g$ bands as $64\times64\times3$ galaxy images first through \textsc{\small SKIRT} \citep{Camps_2020} radiative transfer following the same methodology as in \cite{Faucher_2023}, and then post-processed based on \textsc{\small RealSim} \citep{Bottrell_2017a,Bottrell_2017b,Bottrell_2019} to arrive at realistic mock images.

\textsc{\small SKIRT} is one of the most widely used radiative transfer code to produce idealized synthetic galaxy images from simulations. For each star particle, assuming a Chabrier \citep{Chabrier_2003} initial mass function (IMF), we assign an spectral energy distribution (SED) from FSPS \citep{Conroy_2009,Conroy_2010,Foreman_mackey_2015} using the MIST isochrones \citep{Paxton_2011,Paxton_2013,Paxton_2015,Choi_2016,Dotter_2016} and the MILES spectral library \citep{Sánchez-Blázquez_2006} according to its age, metallicity and mass. Photons are sampled from the SED, launched isotropically in the rest frame of the particle and subsequently Doppler shifted. To reduce the stochasticity of the star-formation histories caused by modeling populations of stars as single particles, we implement a subgrid recipe that effectively smooths out the simulated star formation histories such that the typical difference in age between two neighboring (in the temporal sense) young star particles is less than $\sim1$ Myr, the timescale on which stellar population spectra show significant variation. For star particles younger than 10 Myr, we also need to account for the absorption and emission by dust within photodisassociation regions (PDRs) that result from the remaining birth clouds of newly formed stars. Because these regions are below the spatial resolution of the simulations, we adopt the commonly used method \citep{Groves_2008,Jonsson_2010,Hayward_2015,Trayford_2017,Trčka_2020,Kapoor_2021,Faucher_2023} of assigning SEDs from MAPPINGS-III that already include the effects of photoionization and obscuration within these dense molecular clouds. This model is characterized by a single free parameter which describes the clearing time of molecular clouds and is taken to be 2.5 Myrs following \cite{Faucher_2023}. Because the NIHAO simulations do not directly model the dust population, we assume that each gas particle contains a dust mass given by a 10\% of its mass in metals. We also assume that no dust is present in gas above a maximum temperature of 16,000K. To perform radiative transfer calculations, we discretize space using an OctTree spatial grid that we subdivide until each grid cell contains at most one gas particle. The physical apparent size of each galaxy image is determined by the distribution of dark matter particles belonging to the galaxy’s primary halo, as determined by the AHF halo finder\citep{Knollmann_2009}. The image pixel scale of NIHAO mock images ranges from 0.58 - 3.74 kpc/pixel. More details and example outputs of our radiative transfer procedure can be found in \cite{Faucher_2023}.

We further add observational realism including point spread function (PSF) convolution, shot noise, Gaussian sky noise, and \texttt{arcsinh} stretch based on the \textsc{\small RealSim} code by \cite{Bottrell_2017a,Bottrell_2017b,Bottrell_2019}. To convolve with SDSS PSF, we adopt Gaussian PSF with full width at half-maximum (FWHM) at the average seeing of all SDSS Legacy galaxies ($1.286''$,$1.356''$, and $1.496''$ for SDSS $i$-$r$-$g$ bands). The physical width of the simulated galaxies are converted to angular size by hypothetically putting them at distance of redshift 0.109, the mean redshift of our SDSS training sample. The shot noise is a Poisson noise determined by zeropoints, airmass, extinction, and CCD gain in survey fields. Gaussian sky noise is obtained from the average sky noise over all Legacy galaxies. Lastly, an \texttt{arcsinh} stretch proposed by \cite{Lupton_2004} is implemented to follow SDSS standard imaging scheme.

\section{Observational data}
\label{sec:observation}

\subsection{SDSS galaxy cutouts}
SDSS \citep{2017AJ....154...28B} is one of the largest ongoing surveys to map our Universe. The SDSS cutout tool enables one to get RGB image slices at desired position and width. The red color in SDSS images comes from SDSS near infrared ($i$) filter ($7625 \textup{\AA}$), the green color from SDSS red ($r$) filter ($6231 \textup{\AA}$), and the blue color from SDSS green ($g$) filter ($4770 \textup{\AA}$). To get the galaxy images, we use the galaxy catalog by \cite{10.1093/mnras/stu2333}, which provides the coordinates and stellar masses for 670,722 galaxies. Through the SDSS cutout tool\footnote{We provide the script to download all SDSS dataset at \url{https://github.com/ZehaoJin/Rate-galaxy-simulation-with-Anomaly-Detection/blob/main/SDSS_cutouts/download_cutouts.py}.}, we slice $64\times64$ pixels around each galaxy's coordinate in the pixel scale of the SDSS camera ($0.396''/\textup{pix}$). Examples of SDSS images can be found in Fig.~\ref{fig:overview}. 

We further place a cut on stellar mass at $10^9 \textup{M}_\odot$ to avoid contamination from stars. These SDSS samples have redshift from 0.005 - 0.395, with a mean redshift of 0.109, to be compared with NIHAO snapshots at redshift 0. Finally, these images are split into a training set of 579,197 images, and a test set of 64,356 images. During the training phase, the neural network only sees training set images. After GANomaly is fully trained, the test set will be used to validate the training performance and be used in the analysis presented in this paper.

\subsection{NIHAO vs. SDSS sample statistics}
\begin{figure}
  \centering
  \includegraphics[width=0.85\linewidth]{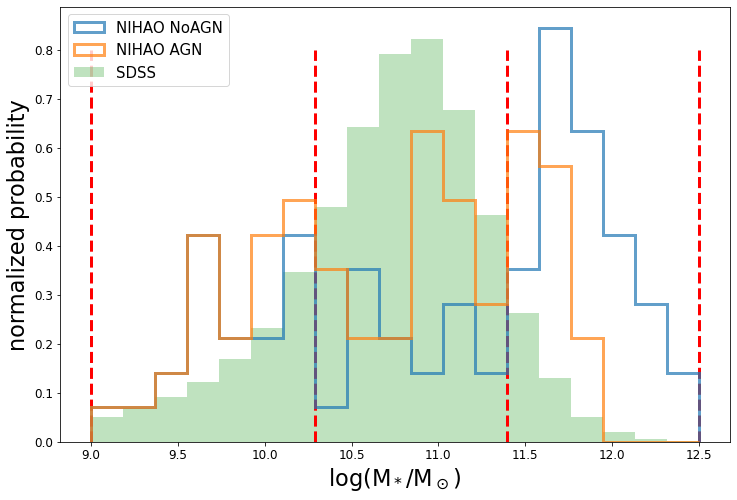}
  \caption{The normalized stellar mass distribution for SDSS (green shade), NIHAO NoAGN (blue line), NIHAO AGN (orange line). The same color scheme will be used throughout this paper. The four red dashed vertical lines further enclose each sample into three (low, middle, high) mass bins. Galaxies in the same mass bin will be compared to each other in the later analysis. It is clear that SDSS's mass distribution peaks around $10^{11} \textup{M}_\odot$, while NIHAO's mass distribution is by construction more flat over all masses, with a slight excess on the higher end.}
\label{fig:stellarmass}
\end{figure}

All NIHAO galaxies presented in this work are selected to have stellar mass $\mathrm{M_*}>10^9 \textup{M}_\odot$, in compliance with the stellar mass cut made on SDSS galaxies\footnote{77 out of 127 pairs of NIHAO NoAGN/AGN galaxies, 12 out of 20 NIHAO n80 galaxies, and 6 out of 6 NIHAO UHD galaxies survives the stellar mass cut.}. Fig.~\ref{fig:stellarmass} shows the stellar mass distribution of NIHAO and SDSS galaxies. All three samples share a similar range in stellar mass, but the exact distribution over the mass range differs. The mass distribution of NIHAO NoAGN and NIHAO AGN is more or less the same, especially at the lower end, since AGN does not play a key role in lower mass galaxies compared to high-mass galaxies. Both NIHAO samples by construction have relatively even distribution among the mass range, with a slightly higher number of higher mass galaxies, but SDSS have a peak in the middle mass bin and deficits in low and high mass bins. Since GANomaly learned purely on SDSS images, it is reasonable that the neural network recognizes medium mass galaxy slightly better than lower mass and high-mass galaxies. We are aware of this selection effect, and we will compare galaxies that are in the same mass bins to overcome this issue. More discussion on how stellar mass effects the anomaly score can be found in Section \ref{sec:results}. NIHAO UHD and NIHAO n80 galaxies will be handled individually later in this paper, since their population is limited. Their stellar mass is very similar to their NIHAO NoAGN counterpart.

\section{GANomaly}
\label{sec:GANomaly}
\subsection{Anomaly detection by reconstruction}
GANomaly \citep{akcay2018ganomaly} is a Generative Adversarial Network (GAN) \citep{NIPS2014_5ca3e9b1} based anomaly detection model inspired by AnoGAN \citep{https://doi.org/10.48550/arxiv.1703.05921}, BiGAN \citep{https://doi.org/10.48550/arxiv.1605.09782}, and EGBAD \citep{https://doi.org/10.48550/arxiv.1802.06222}. GANomaly detects anomaly by image reconstruction. GANomaly is trained to reconstruct normal (non-anomalous) images by learning the commonly shared features in the set of normal images. After training is finalized, GANomaly should only be able to reconstruct normal images, but fail to reconstruct any anomaly. Hence, comparison between original and reconstructed will reveal the anomaly.

\subsection{Network architecture}
\begin{figure*}
  \centering
  \includegraphics[width=0.9\linewidth]{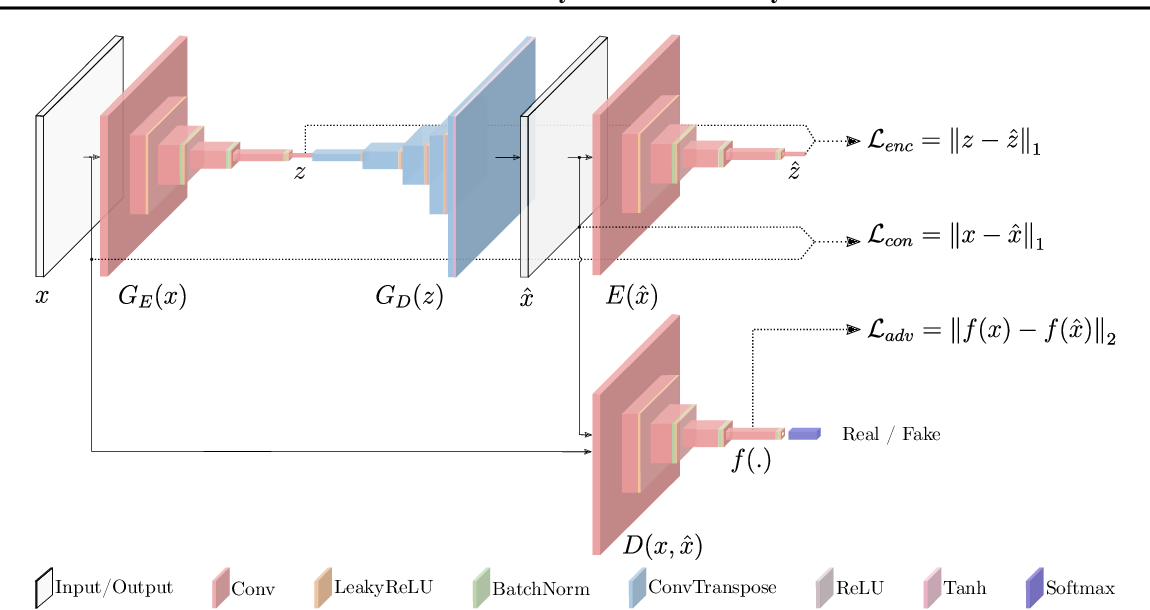}
  \caption{GANomaly architecture, adopted from \protect\cite{akcay2018ganomaly} and \protect\cite{1906.11632}. GANomaly mainly consists of a generator network (encoder $G_{\textup{E}}$ and decoder $G_{\textup{D}}$),  an encoder $E$, and a discriminator network $D$. An input image $x$ first gets encoded by $G_{\textup{E}}$ into $z$, the latent/feature space representation of $x$, and then through $G_{\textup{D}}$ reconstructed back to $\hat{x}$. The reconstructed $\hat{x}$ is further sent to $E$ and encoded into $\hat{z}$, the feature representation of the reconstructed $\hat{x}$. At the same time, the discriminator $D$ anonymously takes both input image $x$ and reconstructed image $\hat{x}$, and trys to distinguish between the two. The training of GANomaly aims to minimize adversarial Loss $\mathcal{L}_{\textup{adv}}$, contextual Loss $\mathcal{L}_{\textup{con}}$, and encoder Loss $\mathcal{L}_{\textup{enc}}$.}
\label{fig:ganomaly}
\end{figure*}

Specifically, Fig.~\ref{fig:ganomaly} visualizes the architecture of GANomaly. As a variation from GAN, GANomaly is made of a generator network (encoder $G_{\textup{E}}$ and decoder $G_{\textup{D}}$) and a discriminator network $D$, with an additional encoder $E$. An input image $x$ ($64\times64\times3$ in this work) first goes through the encoder part of the generator $G_{\textup{E}}$ and being encoded, or summarized as $z$ ($1\times128$ in this work), i.e. the feature space representation of $x$. $z$ then becomes the input of the decoder $G_{\textup{D}}$ that outputs $\hat{x}$ ($64\times64\times3$), the reconstructed version of $x$. Finally, the reconstructed $\hat{x}$ is sent to another encoder $E$ and encoded into $\hat{z}$ ($1\times128$), the feature representation of the reconstructed $\hat{x}$. Meanwhile, the discriminator will anonymously take both original input image $x$ and reconstructed image $\hat{x}$, and try to tell which one is the real input image and which one is the fake image generated by the generator. The generator $G$ and the discriminator $D$ are then in rivalry and grow together: the generator tries to fool the discriminator by generating more and more realistic images, while the discriminator learns to stay sharp and distinguish generated images from the real ones. 

\subsection{Loss and anomaly score}
\label{sec:loss}
To reach the goal of successful image reconstruction, three loss functions are defined and to be minimized during training.

\emph{Adversarial Loss}, $\mathcal{L}_{\textup{adv}}$: The adversarial loss powers the competition between generator and discriminator by telling if the discriminator successfully distinguished the real and generated images. Unlike vanilla GAN where the adversarial loss is simply binary true or false\footnote{A simple binary adversarial loss would work in GANomaly too}, here following \cite{https://doi.org/10.48550/arxiv.1703.05921} and \cite{https://doi.org/10.48550/arxiv.1802.06222} the adversarial loss is based on the internal representation of discriminator $D$. 
\begin{equation}
\mathcal{L}_{\textup{adv}} = ||f(x)-f(\hat{x})||_{2},
\end{equation}
where $f$ is an intermediate layer inside discriminator D. This loss function computes the $\mathcal{L}_{2}$ distance between the feature representation of the original and the generated images. Note that although both being feature representation of $x$, $f(x)$ is different from $z$. $f(x)$ comes from a layer in the discriminator, and the features will be trained to best help distinguish real and generated images; $z$ on the other hand comes from the encoder, and the features will best serve the reconstruction of $x$. 

\emph{Contextual Loss}, $\mathcal{L}_{\textup{con}}$: The contextual loss directly compares the input image and the generated image by an $\mathcal{L}_{1}$ distance,
\begin{equation}
\mathcal{L}_{\textup{con}} = ||x-\hat{x}||_{1}
\end{equation}
Minimizing $\mathcal{L}_{\textup{con}}$ simply pushes input $x$ and constructed $\hat{x}$ to be as identical as possible, thus contextual information of normal images will be learned.

\emph{Encoder Loss}, $\mathcal{L}_{\textup{enc}}$: The encoder loss is the $\mathcal{L}_{2}$ distance between the encoded feature representation of original $x$ and reconstructed $\hat{x}$.
\begin{equation}
\mathcal{L}_{\textup{enc}} = ||z-\hat{z}||_{2}
\end{equation}
Minimizing $\mathcal{L}_{\textup{enc}}$ lets the generator learn how to grasp features of a non-anomalous image.

Overall, the training goal of GANomaly is to minimize the weighted sum of the three losses:
\begin{equation}
\mathcal{L} = w_{\textup{adv}}\mathcal{L}_{\textup{adv}}+w_{\textup{con}}\mathcal{L}_{\textup{con}}+w_{\textup{enc}}\mathcal{L}_{\textup{enc}}
\end{equation}
where $w_{\textup{adv}}$, $w_{\textup{con}}$, and $w_{\textup{enc}}$ enables the adjustment of importance of the three losses.

The anomaly score of an input $x$, $\mathcal{A}(x)$, however, will not use the collective loss function, but only uses the difference in the feature space $\mathcal{L}_{\textup{enc}}$. The contextual loss $\mathcal{L}_{\textup{con}}$, although not directly linked to anomaly score, can be used to infer the location of the anomaly.

\begin{equation}
\mathcal{A}(x) = \mathcal{L}_{\textup{enc}} = ||z-\hat{z}||_{2}
\end{equation}

\subsection{Training}
GANomaly is trained only on the training set of 579,197 real SDSS images. The input/output ($x,\hat{x}$) dimension is set to $64\times64\times3$, and the feature space ($z,\hat{z}$) dimension is chosen to be $1\times128$. Note that the dimension of the feature space is one of the hyper-parameters that are somewhat arbitrary and could be tuned. The `features' are not fully independent and orthogonal to each other, and thus the feature space is hard to interpret. The encoder could summarize an image in 128 bits, but could also fully summarize the same image in 64 bits or 256 bits. The neural network is trained over 50 iterations on the training set, with loss weights $w_{\textup{adv}}=1$, $w_{\textup{con}}=50$, and $w_{\textup{enc}}=1$\footnote{The same loss weight used in \cite{akcay2018ganomaly}. $w_{\textup{con}}=50$ is seemingly high since $\mathcal{L}_{\textup{con}}$ is averaged over $64\times64\times3$ pixels. A local contextual discrepancy will be diluted after averaging, requiring a higher weight.}, batch size 64, learning rate $\alpha=0.0002$, and Adam optimizer $\beta_1=0.5$. The whole training took around 150 hours on a Dual NVIDIA Quadro P1000 GPU with approximately 4GB of memory.

\section{Results}
\label{sec:results}

\subsection{Overview}
\begin{figure*}
  \centering
  \includegraphics[width=0.3\linewidth]{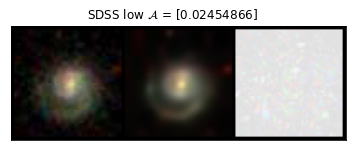}
  \includegraphics[width=0.3\linewidth]{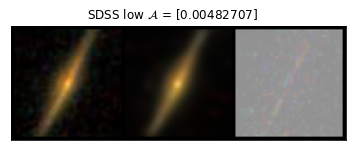}
  \includegraphics[width=0.3\linewidth]{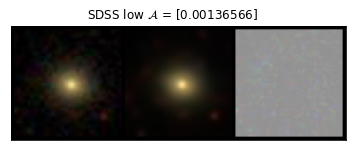}
  \includegraphics[width=0.3\linewidth]{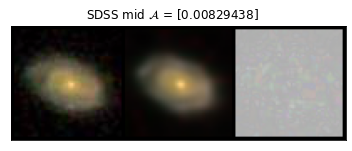}
  \includegraphics[width=0.3\linewidth]{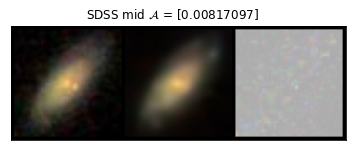}
  \includegraphics[width=0.3\linewidth]{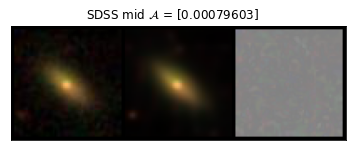}
  \includegraphics[width=0.3\linewidth]{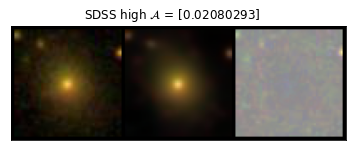}
  \includegraphics[width=0.3\linewidth]{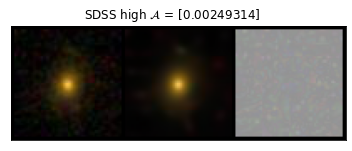}
  \includegraphics[width=0.3\linewidth]{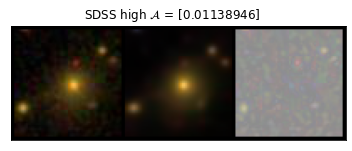}
  \includegraphics[width=0.3\linewidth]{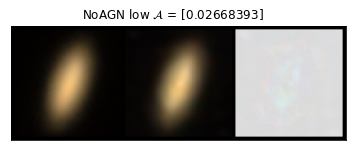}
  \includegraphics[width=0.3\linewidth]{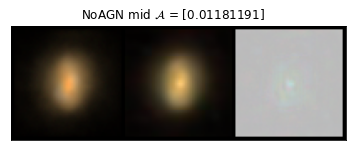}
  \includegraphics[width=0.3\linewidth]{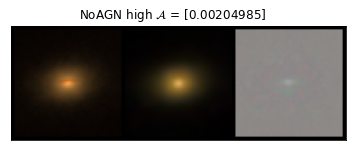}
  \includegraphics[width=0.3\linewidth]{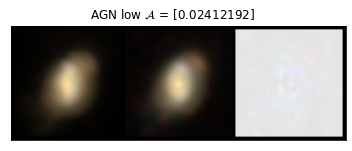}
  \includegraphics[width=0.3\linewidth]{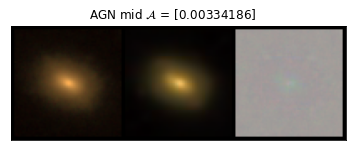}
  \includegraphics[width=0.3\linewidth]{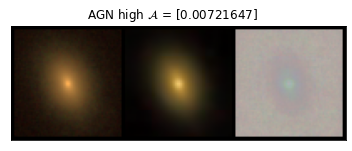}
  \includegraphics[width=0.3\linewidth]{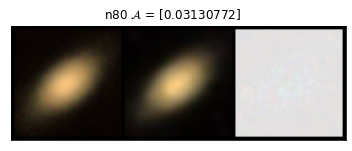}
  \includegraphics[width=0.3\linewidth]{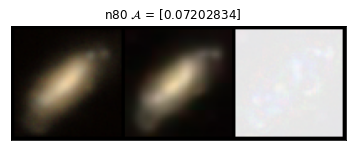}
  \includegraphics[width=0.3\linewidth]{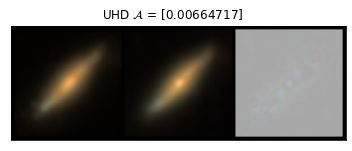}
  \includegraphics[width=0.3\linewidth]{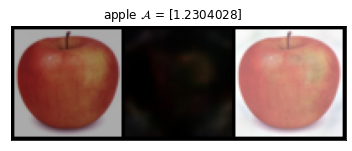}
  \includegraphics[width=0.3\linewidth]{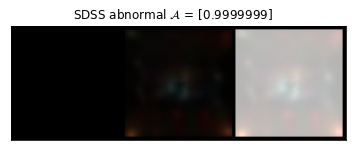}
  \includegraphics[width=0.3\linewidth]{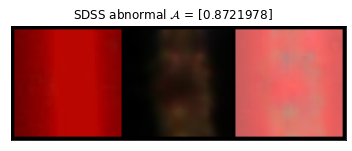}
  \caption{A gallery of GANomaly performance. On each panel of three images, left is the input original image, middle is the GANomaly reconstructed image, right is the residual between input and reconstructed. and on top is the anomaly score. The anomaly score comes from the normalized encoder loss $\mathcal{L}_{\textup{enc}}$, while the residual in the rightmost panel is the pixel-wise contextual loss. A smaller anomaly score indicates the galaxy image is more realistic compared to SDSS galaxy images. The residual on the right hand panels hints at the location of the anomaly, but note that there is no one to one relation between residual and anomaly score (See Section \ref{sec:background}). All images are randomly selected. $1^{st}\sim3^{rd}$ row: SDSS test set galaxy images from low, middle and high mass bins; $4^{th}$ row: NIHAO NoAGN galaxies in each mass bin; $5^{th}$ row: NIHAO AGN galaxies in each mass bin; $6^{th}$ row: NIHAO n80 and NIHAO UHD galaxies; $7^{th}$ row: Sanity check with an apple, and two abnormal SDSS test set images. One when no galaxy is observed and the whole field of view is black, and one when a cosmic ray strikes through the camera.}
\label{fig:overview}
\end{figure*}

\begin{figure}
  \centering
  \includegraphics[width=0.9\linewidth]{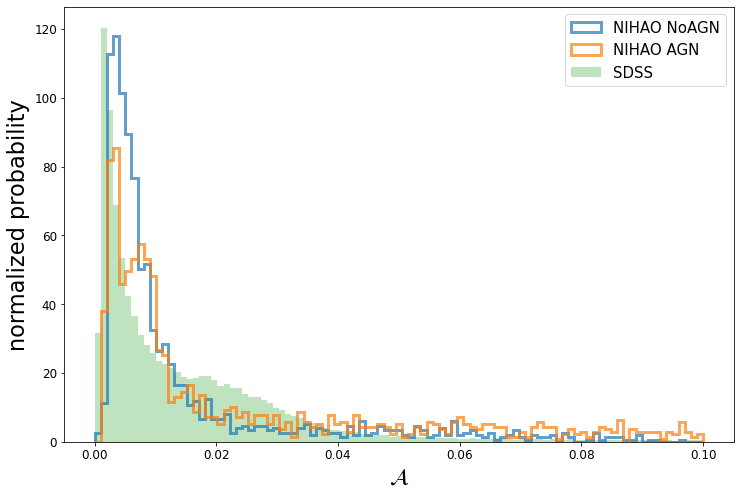}
  \caption{Distribution of anomaly scores for SDSS observations and NIHAO simulations. Each NIHAO galaxy is projected in 20 random orientations, and each of these 20 orientations are seen as independent images when calculating the anomaly score distribution. The distributions do have large overlapping area, but NIHAO does not reproduce the SDSS peak and have a larger tail.}
\label{fig:SDSSnihao}
\end{figure}

After training on the SDSS training set, we apply GANomaly to the SDSS test set and our NIHAO images. Note that the SDSS test set images have never been seen by GANomaly during the training phase, and hereafter all SDSS images mentioned are from the SDSS test set. A gallery of typical GANomaly output is shown in Fig.~\ref{fig:overview} in [original, reconstructed, residual] format, with anomaly score $\mathcal{A}$ on top. Note that the anomaly score is normalized by the lowest and highest scores in the SDSS sample thus ranging between 0 and 1 for SDSS (the highest anomaly score case, 1, is the SDSS `black view' case). A non-SDSS image can get scores higher than 1 if it is even more abnormal than the `black view' case, e.g. the apple in Fig.~\ref{fig:overview}. A lower anomaly score means the image has less anomaly in feature space, or naively, a galaxy image with lower anomaly score is more realistic. As shown in the gallery, SDSS galaxy images are reconstructed nicely, with tiny residuals and very low anomaly score. However, GANomaly fails to reconstruct any anomaly or non-SDSS galaxy, such as an apple, a black view, and a cosmic ray instance (bottom row in Fig.~\ref{fig:overview}). NIHAO simulated galaxies are reconstructed fairly nicely according to the residual map and anomaly score. The anomaly score distribution of all SDSS test set images versus that of all NIHAO NoAGN and NIHAO AGN images is shown in Fig.~\ref{fig:SDSSnihao}. Both NIHAO NoAGN and NIHAO AGN distributions overlap the SDSS distribution by around $70 $ per cent, but both of them are not able to reach the very low score where the SDSS distribution peaks, and both have a larger tail than the SDSS distribution. That indicates NIHAO simulations are generally realistic galaxy simulations, but still have space to improve. Below a more careful interpretation of anomaly scores across different sets of NIHAO simulations will be presented.

\subsection{NIHAO NoAGN vs. NIHAO AGN in mass bins}
\label{sec:NoAGNAGN}
\begin{figure}
  \centering
  \includegraphics[width=0.9\linewidth]{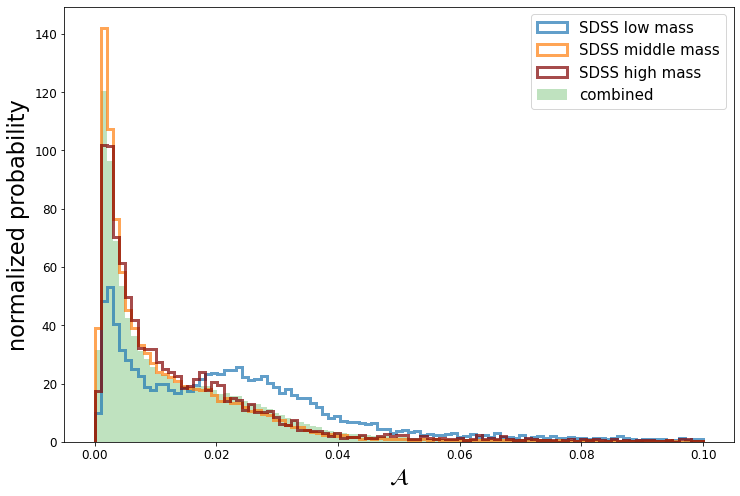}
  \caption{Normalized distribution of anomaly scores for different stellar mass bins in the SDSS test set. Colored lines denote three mass bins and the green shaded histogram shows the combined dataset. The anomaly score favors middle and high-mass SDSS galaxies over lower mass ones.}
\label{fig:SDSSmass}
\end{figure}

Since the stellar mass is not evenly distributed in the SDSS training set (Fig.~\ref{fig:stellarmass}), and different stellar mass can lead to distinct galaxy morphologies, GANomaly will favor galaxies with certain stellar mass, as shown in Fig.~\ref{fig:SDSSmass}. Lower mass galaxies, due to their weaker gravitational potential, are intrinsically more irregular or abnormal in terms of morphology compared to more massive galaxies in middle and high mass bins. The intrinsic anomaly and the unbalanced stellar mass population in training set together make SDSS galaxies in low mass bin  have higher anomaly scores than in mid and high mass bins. For a fair comparison, we put both SDSS and NIHAO galaxies in low, middle, and high mass bins and compare the distribution in the same mass range. 

\begin{figure}
  \centering
  \includegraphics[width=0.9\linewidth]{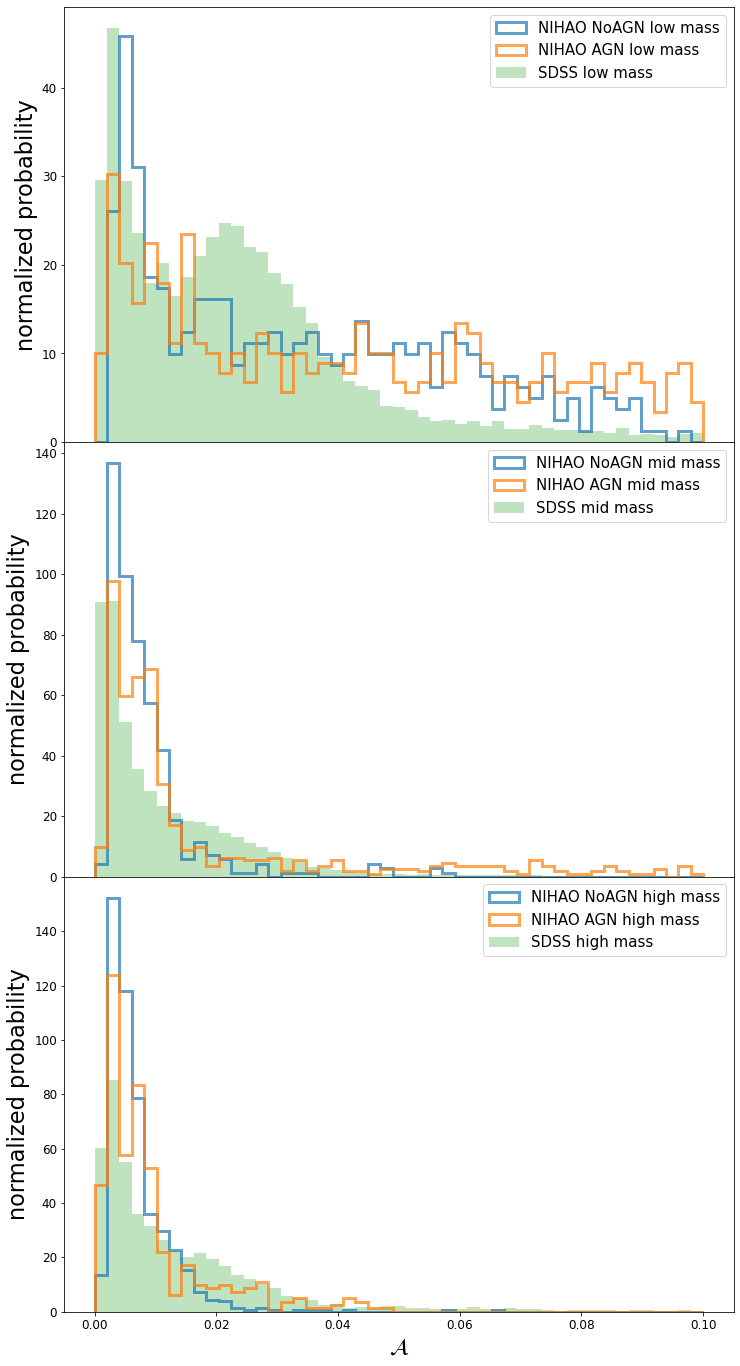}
  \caption{NIHAO NoAGN (blue) vs. NIHAO AGN (orange) vs. SDSS (green) in low, middle and high (top to bottom) mass bins. The 20 projections of a NIHAO galaxy are treated as independent images when calculating the anomaly score distribution. We do not see a significant preference towards NIHAO AGN or NIHAO NoAGN in any of the mass bin.}
\label{fig:NIHAONoAGNAGN}
\end{figure}

AGN feedback is believed to be essential in quenching of massive galaxies, thus one would expect little difference in anomaly score between NIHAO NoAGN and NIHAO AGN in low mass bin, while NIHAO AGN's anomaly score should outperform that of NIHAO NoAGN as we go to higher stellar mass, assuming the AGN implementation is realistic. Fig.~\ref{fig:NIHAONoAGNAGN} shows the anomaly score distribution for NIHAO NoAGN, NIHAO AGN, and SDSS in each of the three mass bins. The overlapping area (in per cent of SDSS area) between NIHAO and SDSS for AGN vs. NoAGN in low, middle, and high mass bins is 62 per cent vs. 68 per cent, 66 per cent vs. 59 per cent, and 74 per cent vs. 63 per cent\footnote{Note that overlapping area is dependent on the choice of binning, therefore does not directly quantify the goodness of NIHAO NoAGN or NIHAO AGN over the other.}. The plot shows that none of NoAGN nor AGN is outperforming the other significantly in all of the mass bins. The overlapping area implies that in the low mass bin, there is little difference between NIHAO NoAGN and NIHAO AGN since black holes do not play a key role there, as expected.  While in the middle mass bin and especially in high mass bin, AGN starts to show slight advantage over NoAGN. Visually, the advantage of AGN in high mass bin comes from that AGN reproduces the second peak around anomaly score of 0.01 - 0.03 a little better than NoAGN does. 

To summarize, GANomaly hints some tiny improvement in the modeling of higher mass galaxies with AGN implementation, but generally draws a tie between with or without AGN on their overall performance. Similarly, using PixelCNN and its log-likelihood ratio (LLR) distribution (similar to anomaly score distribution here) to compare SDSS $r$ band images, \cite{2021MNRAS.501.4359Z} found only a marginal improvement for quiescent galaxies from Illustris to IllustrisTNG despite their distinct AGN feedback implementation. GANomaly takes in mock observed galaxy images, i.e. SDSS $i$-$r$-$g$ band normalized light distribution maps, therefore, current AGN implementations in NIHAO seems to have `no net effect' on normalized galaxy light distribution maps according to GANomaly. In Section \ref{sec:morphology} we will present more on how anomaly scores seem to be `insensitive' to current AGN models in NIHAO.

\subsection{Effect of star formation threshold}
\begin{figure}
  \centering
  \includegraphics[width=0.9\linewidth]{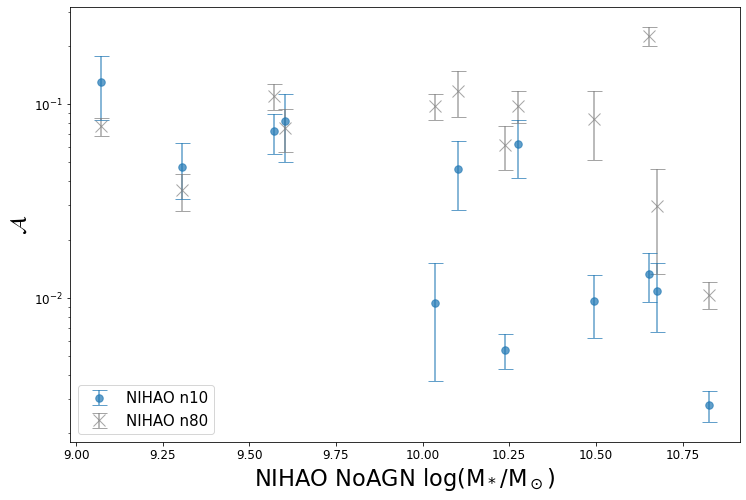}
  \caption{Anomaly score for 12 NIHAO n10 and NIHAO n80 counterparts. Each n10 and n80 galaxy is projected in 20 different rotations, giving 20 galaxy images (see Section \ref{sec:mockobs}). On this plot each dot shows the mean anomaly score of each 20 galaxy images on the y-axis, and the y-error bar denotes the standard deviation of the 20 anomaly scores. The x-axis is the stellar mass of the NIHAO n10 galaxy, just to group each counterpart. Anomaly score favors $n=10\ \textup{cm}^{-3}$ over $n=80\ \textup{cm}^{-3}$ in higher mass galaxies.}
\label{fig:n80score}
\end{figure}

Among 81 NIHAO NoAGN galaxies used in this work, we have 12 galaxies that have NIHAO n80 counterparts. These NIHAO n80 galaxies have no AGN implementation and have star formation density threshold $n=80\ \textup{cm}^{-3}$ instead of $n=10\ \textup{cm}^{-3}$ as in NIHAO NoAGN (See Section \ref{sec:n80}). In this section, we will refer to the comparison between these 12 pairs of galaxies as NIHAO n80 vs. NIHAO n10 for clarity. To investigate this $n$, we compare the anomaly score performance of $n=80\ \textup{cm}^{-3}$ versus $n=10\ \textup{cm}^{-3}$ galaxy by galaxy, as shown in Fig.~\ref{fig:n80score}. For these 12 pairs of galaxies, the ones with lower mass ($<10^{10} \mathrm{M_*}/\textup{M}_\odot$) show no clear difference between n80 and n10 in anomaly score, while the ones with higher mass ($>10^{10} \mathrm{M_*}/\textup{M}_\odot$) show better (lower) anomaly score in NIHAO n10. This implies that $n=10\ \textup{cm}^{-3}$ seems to be a better choice than $n=80\ \textup{cm}^{-3}$ in a NIHAO NoAGN-like setup in high mass galaxies. This is an interesting result that is different from \cite{Macci__2022}, which demonstrates that $n=80\ \textup{cm}^{-3}$ is a better choice based on the investigation of the gas map instead of the stellar light map looked at in this work. The varied outcomes implied from diverse perspectives indicate that one single metric is not enough to tune effective parameters or thresholds in cosmological simulations, and an optimal value of $n$ is worth more study in future work.

\subsection{Effect of resolution}
\label{sec:NoAGNvsUHD}
\begin{figure}
  \centering
  \includegraphics[width=0.9\linewidth]{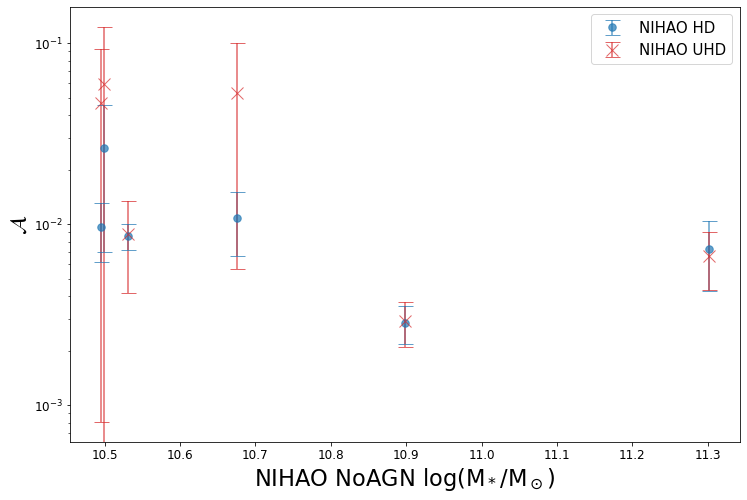}
  \caption{Anomaly score for 6 NIHAO HD and NIHAO UHD counterparts. As in Fig.~\ref{fig:n80score}, each dot shows the mean anomaly score of each 20 galaxy projections on the y-axis, and the y-error bar denotes the standard deviation of the 20 anomaly scores. The x-axis shows the stellar mass of NIHAO HD galaxies to group each counterpart. The anomaly scores for HD and UHD are almost the same, due to the smoothing of PSF.}
\label{fig:UHDscore}
\end{figure}

In a similar fashion as in the n80 case, we compare 6 pairs of NIHAO NoAGN (referred to as `NIHAO HD' in this section) and NIHAO UHD galaxies. As a reminder, the only difference between them is that NIHAO UHD has higher resolution than NIHAO HD. Fig.~\ref{fig:UHDscore} shows that in all 6 galaxies, the anomaly scores are almost identical across different resolution. This is due to that both HD and UHD will effectively have mock images of the same resolution after convolving with the PSF. The same level of anomaly score performance shows that NIHAO HD is able to produce as nice SDSS style galaxy images as NIHAO HUD does at a significantly lower computational cost.

\begin{figure}
  \centering
  \includegraphics[width=0.45\linewidth]{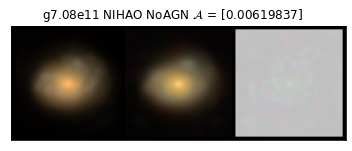}
  \includegraphics[width=0.45\linewidth]{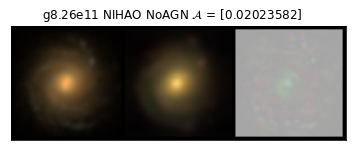}
  \includegraphics[width=0.45\linewidth]{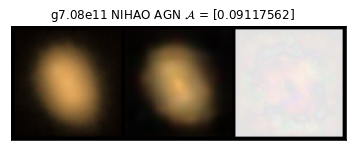}
  \includegraphics[width=0.45\linewidth]{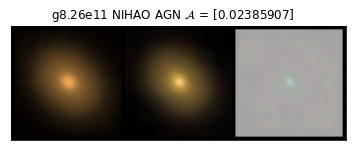}
  \includegraphics[width=0.45\linewidth]{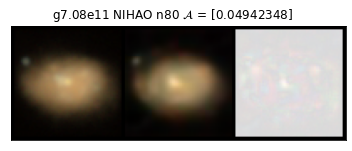}
  \includegraphics[width=0.45\linewidth]{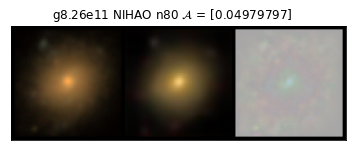}
  \includegraphics[width=0.45\linewidth]{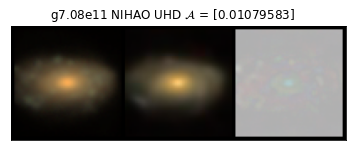}
  \includegraphics[width=0.45\linewidth]{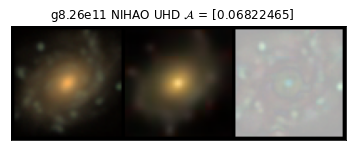}
  \caption{Two NIHAO galaxies named $g7.08e11$ and $g8.26e11$ presented in NoAGN, AGN, n80 and UHD versions. In each set of images left is the original galaxy image, middle is the reconstruction by GANomaly, right is the residual between original and reconstructed, and on top is the name and anomaly score. Note, the projection axes are different across each versions, thus a direct comparison of anomaly scores between these images here is not fair. Besides, the residual in the right hand panels is not necessarily linked to the anomaly score and thus a large reconstruction residual does not explain a large anomaly as explained in section~\ref{sec:loss}.}
\label{fig:gallery2}
\end{figure}

\begin{figure}
  \centering
  \includegraphics[width=0.9\linewidth]{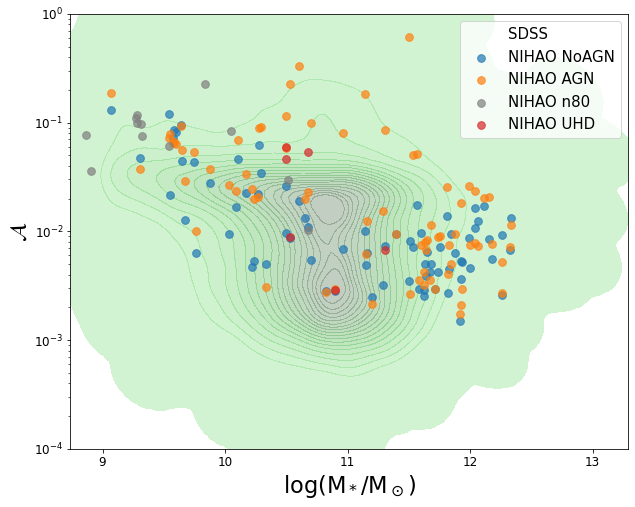}
  \caption{The mean anomaly score for each NIHAO galaxy over its 20 projections, along with the anomaly score distribution for all SDSS test galaxies, plotted as a function of their stellar mass. Green - SDSS; Blue - NIHAO NoAGN; Orange - NIHAO AGN; Grey - NIHAO n80; Red - NIHAO UHD.}
\label{fig:grandscore}
\end{figure}

We also present the face-on galaxy image and GANomaly reconstruction for two NIHAO galaxies that have counterparts in all of NoAGN, AGN, n80 and UHD samples in Fig.~\ref{fig:gallery2}. The mean anomaly score for every NIHAO galaxy and every SDSS test galaxy is visualized in Fig.~\ref{fig:grandscore}. Again in terms of anomaly scores, NIHAO AGN vs. NIHAO AGN draws a tie with marginal advantage for NIHAO AGN at higher masses, NIHAO n10 wins NIHAO n80 (i.e. NIHAO NoAGN) in higher masses, and NIHAO UHD vs. NIHAO HD (i.e. NIHAO NoAGN) draws a tie.

\section{Scaling relations}
\label{sec:scalingrelations}

The compliance to scaling relations is a commonly used criteria to rate simulated galaxies, and it is natural to ask whether the scaling relation criteria agrees with GANomaly anomaly scores. From a scaling relations based point of view, \cite{Arora2023} compares simulated NIHAO galaxies with $\sim 2600$ late-type galaxies from the Mapping Nearby Galaxy at Apache point (MaNGA) survey \citep{Manga2015}. The comparisons are performed using multi-dimensional structural scaling relations using quantities such as size (R), stellar mass ($\mathrm{M_*}$), rotational velocity (V), and stellar surface density within 1 kpc ($\Sigma_1$). Where R, $\mathrm{M_*}$, and $\Sigma_1$ are estimated using optical \textit{grz} photometry from the DESI\footnote{Dark Energy Sky Instrument \citep{DESI}} survey \citep{Arora2021}, while the velocity measurements are \texttt{tanh} model fits to the MaNGA velcity maps \citep[see][for more details]{Arora2023}. For the comparisons all quantities, with the exception of $\Sigma_1$, are measured at a radius corresponding to a stellar mass surface density of $10\,{\rm M_{\odot}\,pc^{-2}}$. The choice of the physically-motivated size metric allows for a uniform comparison between the simulations and observations of galaxies (after considering oberved errors). 

\begin{figure}
  \centering
  \includegraphics[width=0.99\columnwidth]{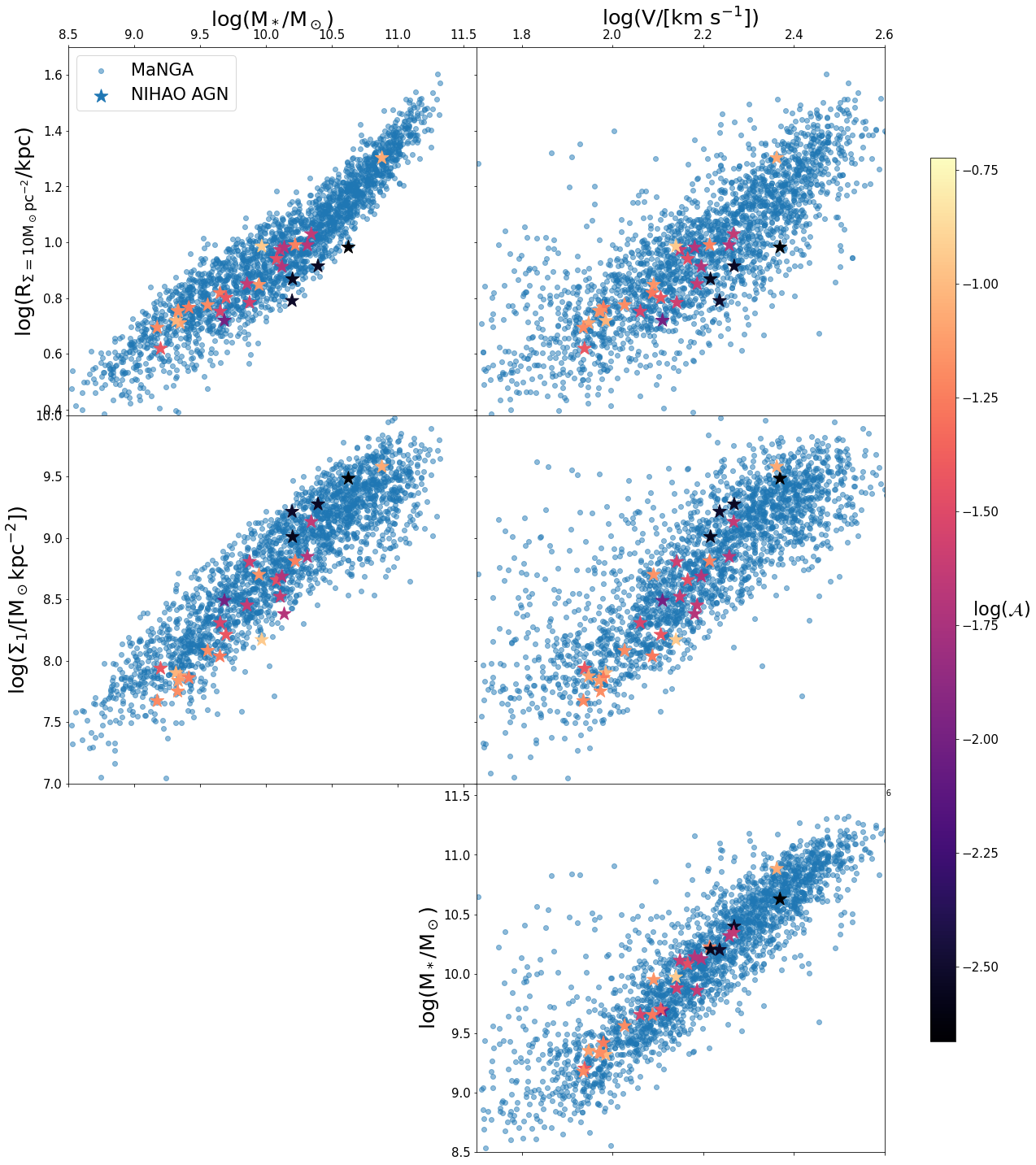}
  \caption{Galaxy scaling relations plotted for NIHAO AGN galaxies, and real galaxies observed in MaNGA. NIHAO AGN galaxies are color-coded by their mean anomaly score over 20 different orientations. All quantities are measured at a radius corresponding to a stellar mass surface density of $10\,{\rm M_{\odot}\,pc^{-2}}$, expect $\Sigma_1$, which is the stellar surface density within 1 kpc. No clear relation between anomaly score and the compliance to scaling relations is found in any of the scaling relation presented.}
\label{fig:scalingrelation}
\end{figure}

In Fig.~\ref{fig:scalingrelation}, we use 30 NIHAO AGN galaxies that are common between the analysis presented here and in \cite{Arora2023}. NIHAO galaxies generally agree with the MaNGA observations well in terms of scaling relations, and none of the galaxies have an exceptionally high anomaly score. However, it turns out that the anomaly score of a galaxy is not correlated with the fact that this galaxy follows any of the scaling relations: a simulated galaxy that lies right on a scaling relation can have higher anomaly score than a galaxy that seems to be off the scaling relation. Naively, one would assume a galaxy that fits closer to a scaling relation is more realistic than a galaxy that is further away, but the anomaly score indicates that this assumption is not always correct. In other words, GANomaly and anomaly score are not parallel, but complementary to scaling relations. GANomaly examines full galaxy images, or light distribution maps and as such the joint distribution of stars position, age, metallicity and mass, which can not be tested through traditional scaling relations. This also suggests that although modern galaxy simulations can reproduce many observed scaling relations, such success in a few statistical quantities does not guarantee a realistic galaxy image. To rate the quality of a galaxy simulation, getting the light map correct could be as crucial as to get traditional scaling relations right. As we pursue increasingly more precise simulations and a more complete picture of galaxy formation models, it is necessary to make use of both traditional scaling relations and deep learning driven image anomaly detection techniques to get a better understanding of simulations and thus to learn a more complete picture of our Universe.

\section{Morphological parameters}
\label{sec:morphology}
Scaling relations are complementary to anomaly scores, but morphological parameters are expected to be more in alignment with anomaly scores, as they both come from galaxy images. Here we investigate what the machine learning model may learn in addition to the traditional methods applied to in studies of mock images so far. Here we are mainly looking at the Gini-$M_{20}$ \citep[gini coefficient, $M_{20}$, and bulge statistics,][]{Lotz_2004}, CAS statistics \citep[concentration, asymmetry, and smoothness,][]{Conselice_2003}, and MID statistics \citep[multimode, intensity, and deviation,][]{Freeman_2013}. A nice review of the definition of these parameters can be found in \cite{Rodriguez-Gomez_2019}. We calculate morphological parameters for NIHAO NoAGN, NIHAO AGN, and SDSS test set galaxy images with the \textsc{\small statmorph} code outlined also in \cite{Rodriguez-Gomez_2019}. 

\begin{figure*}
  \centering
  \includegraphics[width=0.99\linewidth]{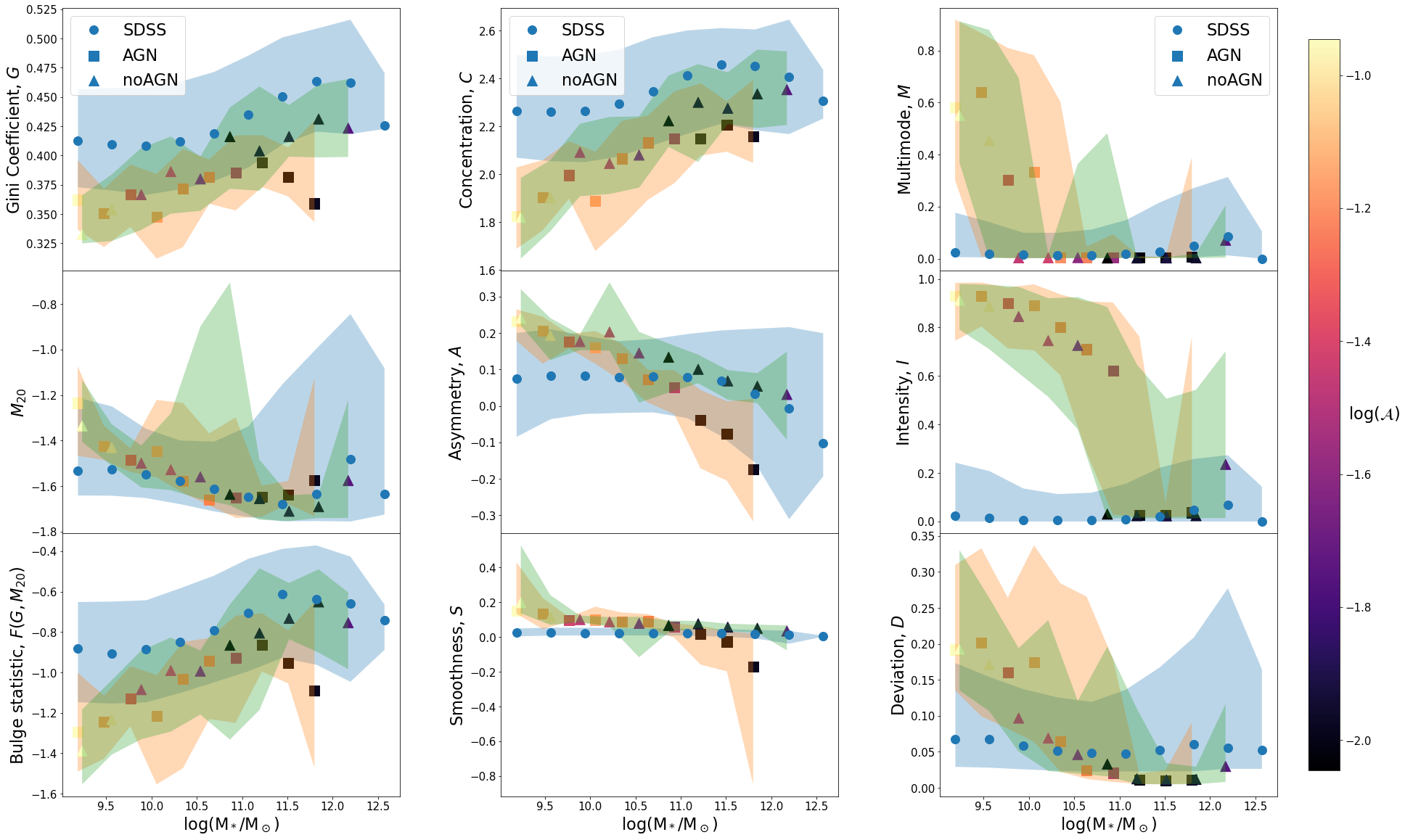}
  \caption{Various morphological parameters compared to anomaly scores for SDSS (circle), NIHAO AGN (square), and NIHAO NoAGN (triangle). The y-value of the points shows the median trend of morphological parameters, and the shaded region indicates the $16^{\mathrm{th}}$ to $84^{\mathrm{th}}$ percentile range. NIHAO galaxies are color-coded by their mean anomaly score over 20 different orientations. In general, NIHAO galaxies have a lower (darker color) anomaly score when the morphological parameter matches that of SDSS better, and this is most obvious in the right column of MID statistics. However, anomaly score and morphological parameters do not always agree, such as in asymmetry, smoothness, and $M_{20}$.}
\label{fig:morphology}
\end{figure*}

In Figure \ref{fig:morphology} we compare anomaly scores to morphological parameters across NIHAO an SDSS. Many of the morphological parameters agree with anomaly score, such as in MID statistics. In such cases NIHAO galaxies have a better anomaly score when the distribution of morphological parameters matches that of SDSS galaxies' better. Further more, the general performance of NIHAO NoAGN and NIHAO AGN are quite similar in terms of most morphological parameters, this is in agreement with what is found in Section \ref{sec:NoAGNAGN} based on anomaly scores. However, anomaly scores seem to disagree with some other morphological parameters. For example, in the case of asymmetry and smoothness\footnote{The high asymmetry and smoothness values at the low-mass end of the galaxy population could be potentially caused by the choice of single Gaussian PSF instead of the non-Gaussian PSF in real SDSS images\citep{Stoughton_2002,Xin_2018}, since the clumpy star-forming regions are often compact in size, as shown in \cite{Bignone_2020,deGraaff_2022}. It is possible that the increased asymmetry value can also effect anomaly score, which is a collective metric that consists of 128 features. We will explore the weights between different features in a further work.}, the lowest anomaly scores for NIHAO AGN are found in highest masses, where the distribution of asymmetry and smoothness between NIHAO AGN and SDSS diverges. Besides, in the case of $M_{20}$, although the distribution of $M_{20}$ between NIHAO and SDSS matches fairly well across all mass bins, the anomaly score can still vary. The agreement between morphological parameters and anomaly scores shows that there are some overlaps between these the approaches, while the disagreement between morphological parameters and anomaly scores suggests that the two approaches are not equivalent. 

Unlike scaling relations that are based on physical properties, morphological parameters and anomaly scores are both based on galaxy images. It is then not surprising that scaling relations and anomaly scores are complementary, while morphological parameters and anomaly scores align more closely, sharing many overlaps. However, morphological parameters are supervised metrics to characterize galaxy images, while GANomaly is instead an unsupervised approach that summarizes a galaxy image into 128 features. The supervised approach by construction guarantees more interpretability, while the unsupervised approach aims to exploit the full information of galaxy images as much as possible. GANomaly, which is optimized to extract features that can fully reconstruct a SDSS-style galaxy image back, can potentially exploit more information from a galaxy image than a set of human-defined morphological parameters. The exact connection between anomaly score and morphological parameters can be explored by correlating the GANomaly feature space and morphological parameters. However, the current GANomaly feature space size of 128 is too large to be easily interpreted. In the Appendix \ref{sec:latent} we attempted to bring the dimension of feature space down by a principal component analysis (PCA). In a upcoming work, we are introducing sparsity into the GANomaly feature space and encourage the feature space to automatically shrink into an optimal size. The sparse-GANomaly feature space will be more interpretable, and their connection to morphological parameters will be studied thoroughly, allowing the full exploitation of mock images while maximizing interpretability.

\section{Background}
\label{sec:background}
\begin{figure}
  \centering
  \includegraphics[width=0.66\linewidth]{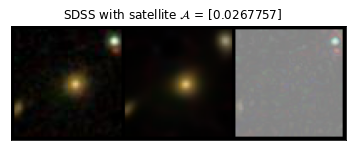}
  \includegraphics[width=0.66\linewidth]{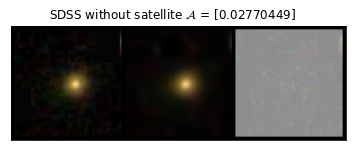}
  \caption{The top row shows a SDSS galaxy image that has some satellites in its background, the bottom row shows the same SDSS galaxy but has the satellites manually removed. Different background results in different residual (right most panel, contextual loss), but almost the same anomaly score (labelled on top, encoder loss).}
\label{fig:satellite}
\end{figure}

Staring at any `galaxy image' referred in this work (e.g. Fig.~\ref{fig:overview},\ref{fig:gallery2}), one might notice that a large amount of image area is occupied by background (black spaces and satellites), but not the galaxy of interest. \cite{2021MNRAS.501.4359Z} pointed out that background has a not negligible impact on the anomaly judgement on one single PixelCNN. PixelCNN works analogous to GANomaly's contextual loss $\mathcal{L}_{\textup{con}}$, in which each the original and reconstructed image is compared pixel by pixel. Any difference in any pixel will be reflected in the anomaly score, regardless of whether the pixel belongs to the galaxy or background. \cite{2021MNRAS.501.4359Z} had to resolve this issue by training two separate PixelCNN. However, in GANomaly, anomaly score is defined only by encoder loss $\mathcal{L}_{\textup{enc}}$, the difference between original and reconstructed features. Any noise in the background that is not a common feature among the training set should not significantly effect anomaly score. For example, in Fig.~\ref{fig:satellite}, a randomly chosen SDSS galaxy has its background satellite manually removed. Such change in background does result in an obvious difference in the pixel-wise residual, as predicted by \citep{2021MNRAS.501.4359Z}, but the anomaly score stays rather stable.

\section{Conclusions}
\label{sec:conclusion}
In this paper, we introduced an anomaly detection algorithm GANomaly that is trained only on real galaxies to quantitatively rate galaxy simulations. Building on the idea of anomaly detection by reconstruction, GANomaly is a combination of a GAN and an autoencoder network to first summarize an input image into its feature/latent space representation, and then reconstruct the same images back. GANomaly is purely trained on normal set of data, therefore once trained, any outlier images or anomalies on the image will not be reconstructed. The anomaly is quantified by defining the anomaly score as the difference in feature space before and after reconstruction. Furthermore, we are only interested in relative anomaly scores between different sets of simulations. Note, this strategy further ensures that any inconsistency in the process of mock image generation will not enter.

To rate galaxy simulations against real observations, we treat SDSS $i$-$r$-$g$ band images as normal set data to train GANomaly, and then apply the trained model to rate mock observations of NIHAO galaxy simulations. Comparing NIHAO simulations with or without AGN implementation, we find that the current AGN model in NIHAO does not improve nor undermine the overall quality of galaxy images. Besides, we find extra resolution in simulation does not effect the quality of a mock image after convolution.

More importantly, we also see that the compliance to galaxy scaling relations does not directly correlate with anomaly scores. This suggests that the success in reproducing certain sets of galaxy properties and galaxy scaling relations does not sign the ultimate victory in simulating our Universe, but a simulated galaxy that fits scaling relations well can still show anomalies when looking at the full galaxy image. Similarly, looking at the images alone might miss some important physical insights that can not be fully described by the image itself. To achieve the ultimate goal of modeling our Universe, simulations need to reproduce both observed scaling relations and realistic galaxy images.

On the other hand, both morphological parameters and GANomaly are devoted to probe galaxy images. Morphological parameters examine image in a supervised way, with clearly defined parameters, thus more interpretable. While GANomaly is an unsupervised deep learning method that builds up a feature space from data, such that the feature space can fully represent an realistic galaxy image. Both two approaches clearly have their own advantages, and combining both will allows us to harness the power of deep learning without losing interpretability, thus fully exploit the rich information carried by galaxy images.

The GANomaly model described in this paper examines galaxy images observed, or mock observed in SDSS $i$-$r$-$g$ band. The model is never restricted to any particular suite of simulations such as NIHAO, but can be directly applied to any other galaxy simulation once SDSS $i$-$r$-$g$ band mock observations are made. The algorithm itself is not limited to a particular telescope or survey (SDSS), or a certain set of maps ($i$-$r$-$g$ band intensity map). GANomaly can be re-trained on images from other telescopes, with different wavelengths, or even different maps, such as velocity maps, and density maps, to detect anomalies either in simulated data, or outliers in the observations themselves.

As we are heading towards an era of next-generation simulations with more sophisticated models, higher resolution, larger volumes, more physics included, and at the same time, a bursting age of big observational missions and large surveys, analyzing techniques that can make the best use of this huge data feed (beyond pure summary statistics) from both simulations and observations is highly needed. Anomaly detection techniques, like GANomaly, along with traditional methods like scaling relations and morphological parameters, together will shed some fresh light on our understanding of the Universe.

\section*{Acknowledgements}
This material is based upon work supported by Tamkeen under the NYU Abu Dhabi Research Institute grant CASS. This research was carried out on the High Performance Computing resources at New York University Abu Dhabi. We gratefully acknowledge the Gauss Centre for Supercomputing e.V. (www.gauss-centre.eu) for funding this project by providing computing time on the GCS Supercomputer SuperMUC at Leibniz Supercomputing Centre (www.lrz.de). The research was made possible by many incredible \textsc{\small Python} packages, including \textsc{\small Pynbody} \citep{pynbody}, \textsc{\small PyTorch} \citep{pytorch}, \textsc{\small Astropy} \citep{astropy:2013, astropy:2018, astropy:2022}, \textsc{\small NumPy} \citep{harris2020array}, \textsc{\small Matplotlib} \citep{Hunter:2007}, and \textsc{\small seaborn} \citep{Waskom2021}. MP acknowledges financial support from the European Union’s Horizon 2020 research and innovation programme under the Marie Sklodowska-Curie grant agreement No. 896248. TB's contribution to this project was made possible by funding from the Carl Zeiss Foundation. 

\section*{Data Availability}

The ready-to-use GANomaly model to rate galaxy simulations, along with the code to produce all the plots for this paper, is located at \url{https://github.com/ZehaoJin/Rate-galaxy-simulation-with-Anomaly-Detection}.

\bibliographystyle{mnras}
\bibliography{bibliography} %

\appendix
\section{Exploring the GANomaly latent space}
\label{sec:latent}
To gain a better understanding of the latent variables used by GANomaly to encode our galaxy images, we run a principal component analysis (PCA) on the 128-dimensional latent space \emph{z}. Note that anomaly score is defined by the distance in latent space before and after reconstruction, $z$-$\hat{z}$. Here a principle component found to be important in $z$ only means the principle component is significant for the image pre-reconstruction, but does not reveal anything on the anomaly score, $z$-$\hat{z}$. Therefore here, we are not exploring what anomaly score means nor what is the criteria for GANomaly to rate a galaxy image, but are trying to understand what are the most important properties for a galaxy image.

\begin{figure}
  \centering
  \includegraphics[width=0.9\linewidth]{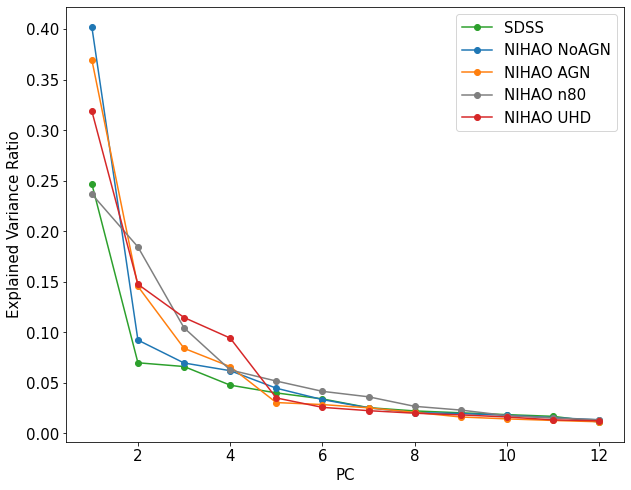}
  \includegraphics[width=0.9\linewidth]{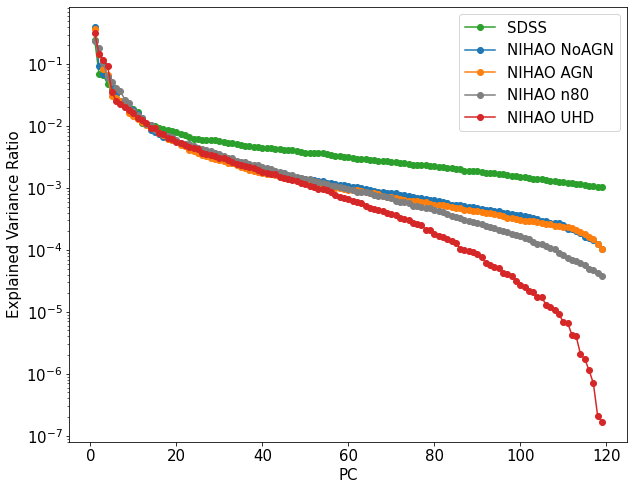}
  \caption{Fraction of explained variance as a function of principal component rank. SDSS is shown in brown and the NIHAO samples in different shades of blue (NIHAO No AGN in light blue, NIHAO AGN in purplish blue, NIHAO UHD in dark blue, NIHAO n80 in greenish blue). The top panel shows only the first twelve components with the y-axis in linear scale and the bottom panel shows all 128 components with the y-axis in log scale.}
\label{fig:PCA}
\end{figure}

In Fig.~\ref{fig:PCA}, we show the fraction of explained variance of the first twelve principal components (top) and of all principal components in log scale (bottom). The PCA was carried out separately for the SDSS galaxies and for each group of NIHAO simulations. It is clear from these plots that the intrinsic dimensionality of the latent space is relatively high, because the first principal component for SDSS samples explains only about $25$ per cent of the total variance. While the general behavior of SDSS and the various NIHAO samples is similar, there is a difference in the fact that the least important components explain more variance in SDSS with respect to NIHAO. We speculate that this may be due to actual observational noise from instrumental and other effects being hard to compress.

\begin{figure}
  \centering
  \includegraphics[width=0.9\linewidth]{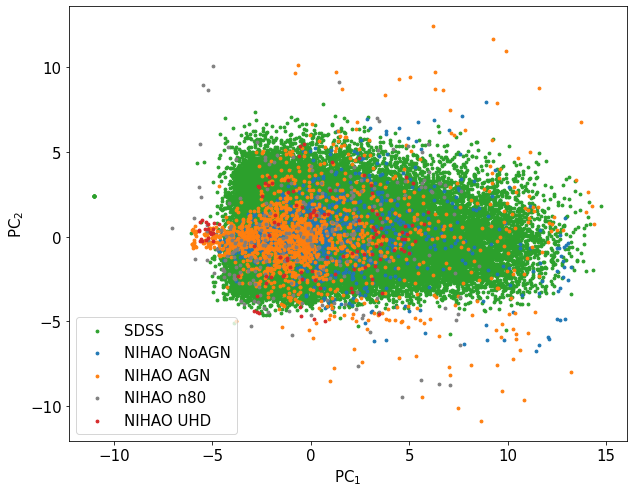}
  \caption{Plot of the first against the second principal component for SDSS galaxies (green). The NIHAO galaxy samples have been projected on the first two SDSS principal components and are shown in various shades of blue.}
\label{fig:PCA_12}
\end{figure}

Apart from this, we find that the NIHAO samples generally fall within the realistic range of latent space spanned by the SDSS galaxies, as shown in Fig.~\ref{fig:PCA_12}, where we plot them in the plane of the first two principal components. Some of the NIHAO galaxies have slightly lower PC1s than SDSS galaxies have, and some of the NIHAO galaxies have higher or lower PC2s than SDSS galaxies do.

\begin{figure}
  \centering
  \begin{tabular}{c}
  \includegraphics[width=0.9\linewidth]{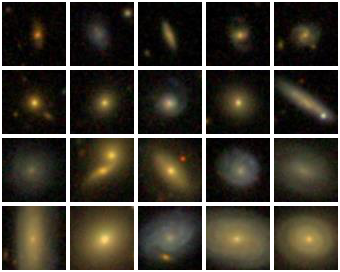}\\
  \phantom{dude} \\
  \includegraphics[width=0.9\linewidth]{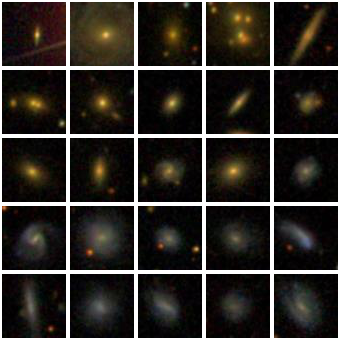}\\
 \end{tabular}
  \caption{Images of a sample of SDSS galaxies ordered by the first principal component (top panels) and by the second principal component (bottom panels).}
\label{fig:PCA_images}
\end{figure}

In Fig.~\ref{fig:PCA_images} we show a sample of SDSS galaxy images ordered by the value of the first PC (top panel) and a sample ordered by the value of the second PC (bottom panel). The first principal component appears to correlate with the apparent size of the galaxy in the image, and the second with color. In the light of this interpretation, Fig.~\ref{fig:PCA_12} shows that some of the NIHAO galaxies have apparent size that are too small, which can be addressed in future work. Some other NIHAO galaxies have too extreme colors which can due to several reasons like e.g. a difference in SFR or the limitations of the dust model but the exact physical meaning of this needs to be explored in more detail and we leave it for future work.

\bsp	%
\label{lastpage}
\end{document}